\documentclass[conference,a4paper]{IEEEtran}

\usepackage{amsmath,amssymb,amsmath,amsfonts}
\usepackage{bm}
\usepackage{bbm}
\usepackage{graphicx}
\usepackage{cite}
\usepackage{epstopdf}
\usepackage{gensymb}
\usepackage{color}
\usepackage{lipsum}
\usepackage{float}
\usepackage{epstopdf}
\usepackage[mathcal]{eucal}
\usepackage{subfiles}
\usepackage[T1]{fontenc}
\usepackage{siunitx}
\usepackage[hyphens]{url}
\usepackage[breaklinks=true]{hyperref}
\usepackage{tabulary}
\usepackage{epsfig,graphics,graphicx,subfig,amssymb,amstext,amsmath,algorithm,algorithmic,wrapfig,multirow}
\usepackage{array}
\usepackage{adjustbox}
\usepackage[dvipsnames]{xcolor}
\usepackage{cite}
\usepackage{times}
\usepackage{placeins}
\usepackage{textcomp}
\usepackage[font=small]{caption}
\usepackage{breakurl}
\usepackage{flushend}
\usepackage{color, colortbl}
\usepackage{tabularx}
\usepackage{bm}
\usepackage{enumerate}
\usepackage{tikz}
\usepackage{pgfplots}
\usepackage{epstopdf}
\usetikzlibrary{shapes,arrows}
\usepackage[utf8]{inputenc}
\usepackage{mathtools}
\usepackage[utf8]{inputenc}
\usepackage{xcolor}
\usetikzlibrary{chains,arrows,calc,positioning}
\usepackage{amssymb}
\usepackage{pifont}
\usepackage{soul}
\usepackage{breqn} 
\usepackage{booktabs} 
\usepackage{multirow}
\usepackage{enumitem}
\usepackage{tabulary}
\usepackage[normalem]{ulem}
\usepackage{dblfloatfix}
\usepackage{makecell}
\usepackage{lipsum}
\usepackage{amsthm}
\usepackage{comment}
\usepackage{ifthen}
\usepackage{xcolor}

\newtheoremstyle{mystyle}
  {}
  {}
  {\itshape}
  {}
  {\bfseries}
  {.}
  { }
  {}
\theoremstyle{mystyle}

\newlength \figwidth
\pgfplotsset{compat=1.16}
\definecolor{bittersweet}{rgb}{1.0, 0.44, 0.37}
\definecolor{glaucous}{rgb}{0.38, 0.51, 0.71}
\definecolor{gainsboro}{rgb}{0.86, 0.86, 0.86}
\definecolor{babyblueeyes}{rgb}{0.63, 0.79, 0.95}
\definecolor{silver}{rgb}{0.75, 0.75, 0.75}
\definecolor{neoncarrot}{rgb}{1.0, 0.64, 0.26}
\definecolor{Gray}{gray}{0.6}
\definecolor{LightCyan}{rgb}{0.88,1,1}
\definecolor{BackgroundLightBlue}{rgb}{0.97,0.97,1}
\definecolor{BackgroundGray}{gray}{0.98}

\newcommand{\blue}[1]{\textcolor{black}{#1}}

\definecolor{boristext}{rgb}{0.22, 0.44, 0.88}
\definecolor{boriscomments}{rgb}{0.88, 0.04, 0.04}
\definecolor{boristochange}{rgb}{0.2, 0.8, 0.8}

\makeatletter

 \let\oldforeign@language\foreign@language
 \DeclareRobustCommand{\foreign@language}[1]{%
   \lowercase{\oldforeign@language{#1}}}

\def\nb0{{\mathbf{0}}}
\def\nb1{{\mathbf{1}}}

\makeatother

\IEEEoverridecommandlockouts

\begin{document}
\title{What Will Wi-Fi 8 Be? A Primer on\\IEEE 802.11bn Ultra High Reliability}

\author{
\IEEEauthorblockN{Lorenzo Galati-Giordano, Giovanni Geraci, Marc Carrascosa, and Boris Bellalta}
}

\bstctlcite{IEEEexample:BSTcontrol}
\maketitle
%
\begin{abstract}

What will Wi-Fi 8 be? 
Driven by the strict requirements of emerging applications, next-generation Wi-Fi is set to prioritize Ultra High Reliability (UHR) above all. 
In this paper, we explore the journey towards IEEE 802.11bn UHR, the amendment that will form the basis of Wi-Fi\,8. 
We first present new use cases calling for further Wi-Fi evolution and \blue{associated standardization, certification, and spectrum allocation efforts.}
We then introduce \blue{a selection of the main disruptive features envisioned for Wi-Fi\,8 and their associated research challenges, resulting from the outcome of the UHR Study Group.} 
Among those, we focus on multi access point coordination and demonstrate that it could build upon 802.11be multi-link operation to make \blue{UHR} a reality in Wi-Fi\,8.
\end{abstract}
\section{Introduction} \label{sec:Intro}

You do not need to be tech-savvy to know Wi-Fi. With twice as many devices as people, Wi-Fi technologies carry two thirds of the world’s mobile traffic and underpin our digital economy. This generation will not easily forget what it could have meant to undergo Covid lockdown without Wi-Fi from social, economic, and safety standpoints. And even now that traveling to places is possible once again, many of us reach for the Wi-Fi password first thing upon arrival, as this is often the means to ordering a meal and sending news back home.

Wi-Fi has come a long way since its introduction in the late nineties. The easiest way to appreciate the technology’s improvement is by reading peak data rates specifications on commercial Wi-Fi access point (AP) boxes. These rates have grown roughly four orders of magnitudes in two and a half decades, from the mere 1\,Mbps of the original 802.11 standard to the near 30\,Gbps of the latest 802.11be products (alias Wi-Fi\,7) scheduled to hit the shelves as early as 2024 \cite{lopez2019ieee,garcia2021ieee,khorov2020current,CheCheDas22}.
This giant leap allowed Wi-Fi to move beyond email and web browsing and progressively conquer crowded co-working spaces, airports, and even the hearts of many parents who can now video-call their children without worrying about phone bills. But how many of us have complained at least once about Wi-Fi not functioning when we most need it? Unreliability would be the Achilles heel for any technology meant to be affordable, pervasive, and operating in license-exempt bands subject to uncontrolled interference. Wi-Fi is no exception.

And while it only takes patience to cope with a buffering video or to repeat our last sentence in a voice call, a lack of Wi-Fi reliability will not be tolerated by its new users: machines. 
In future manufacturing environments, Gbps communications between robots, sensors, and industrial machinery will demand reliability---with at least three (but sometimes many more) ‘nines’---in terms of both data delivery and maximum latency. Rest assured that these requirements will not get any looser for use cases involving humans. Many of us may not even want to think about undergoing robotic-assisted surgery with an unreliable Wi-Fi connection. But even just for holographic communications, a key building block of the upcoming Metaverse, excessive delays experienced by just 0.01\% of the packets could trigger nausea and user distress. As it takes up ever more challenging endeavors to fuel industrial automation, digital twinning, and tele-presence, next-generation Wi-Fi is bound to step out of its comfort zone and set reliability as its first priority~\cite{UHRProposedPAR,ResCor22}.

In this paper, we embark on a journey towards 802.11bn Ultra High Reliability (UHR), the amendment that will form the basis of Wi-Fi\,8. 
\blue{After presenting the emerging applications that are driving a further Wi-Fi evolution, we review the current activities in terms of standardization, certification, and spectrum allocation, and provide a digested summary of the main outcomes produced by the UHR Study Group.} As the research community shifts gears to target new use cases and requirements, \blue{we introduce some of the new features that Wi-Fi\,8 may bring about}, along with their associated research challenges. Among these features, we highlight the multi-AP coordination framework as a game-changer for Wi-Fi, boosting spectrum utilization efficiency and closing in on performance determinism. We also present novel results demonstrating how such disruptive enhancements could build upon 802.11be multi-link operation \blue{(MLO) to maximize their impact, bringing Wi-Fi\,8---and its ultra-reliability ambitions---one step closer}.
%
%
\begin{table*}[h]
\vspace{1mm}
\caption{Representative emerging use cases for Wi-Fi beyond 2030 and their corresponding requirements \cite{UHRProposedPAR,HexaD13}.}
\label{tab:UseCases}
\centering
\colorbox{BackgroundGray}{%
\begin{tabular}{ |m{5.0cm}|m{1.8cm}|m{1.8cm}|m{3.2cm}|m{4.0cm}| } 
\toprule
\rowcolor{BackgroundLightBlue}
 \textbf{Use case} & \textbf{Reliability} & \textbf{Latency} & \textbf{Data rate} & \textbf{Notes} \\ \midrule
\textbf{Immersive communications:} \newline
{Physically present and holographically telepresent consumers with AR glasses and body sensors/actuators.} 
& 
99.9\%
&
<\,20\,ms
&
1-10\,Gbps downlink (DL) \newline (AR stream + spatial map)
\newline
0.1\,Gbps uplink (UL) \newline (spatial map + user data)
&
End-to-end roundtrip UL+DL <\,100\,ms
for 99.99\% of packets to avoid nausea and user distress and
discomfort.
\\ \midrule
\textbf{Digital twins for manufacturing:} \newline
{Machine-to-machine traffic between sensors, alarms, fixed machinery, and moving autonomous vehicles. Human traffic from operators monitoring the factory.}
& 
99.9\% \newline to \newline 99.999999\%
&
0.1\,--\,100\,ms
&
1\,--\,10\,Gbps (average)
\newline to \newline
10\,--\,100\,Gbps (peak)
&
Lower reliability for process and
asset monitoring and higher reliability for
motion control and alarms.
\\ \midrule
\textbf{e-Health for all:} \newline
{Connectivity for users (healthcare professionals, patients, administrators) and devices (smart medical instruments, wearables, on- and in-body sensors/actuators).}
& 
99.999\% \newline to \newline 99.9999999\% &
0.1\,--\,100\,ms
&
100\,kbps (sensor data) \newline to  \newline 25\,Mbps (4K video)
&
Lower latency required in robotic-assisted surgery operations. Peak bit rates can be much higher for specific applications, e.g., XR remote diagnostics.
\\ \midrule
\textbf{Cooperative mobile robots:} \newline
{Communication between robots and static machinery. Direct (through XR devices) or intent-based (trajectory crossing) human-machine interaction on a shop floor.}
& 
up to \newline 99.9999999\%
&
0.5\,--\,25\,ms
&
< 0.1\,Mbps (for control)
&
Deterministic communication required for control applications.
\\ \bottomrule 
\end{tabular}
}
\end{table*}
%
\blue{\section{Emerging Use Cases Driving Novel Standardization Efforts}} \label{sec:updateWiFi}
As Wi-Fi keeps evolving, new use cases and applications are emerging that require not only more throughput and less delay, as targeted by Wi-Fi\,7, but also a higher reliability. In this section, we explore some of the emerging use cases for Wi-Fi\,8, as well as the standardization and regulatory activities that are shaping its development.

\subsection{Emerging Applications and Use cases} \label{sec:usecases}

The key use cases in 2030 and beyond for indoor connectivity 
are foreseen to include the following \cite{UHRProposedPAR,HexaD13}. 

\noindent\emph{Immersive communications}: Moving from augmented/virtual reality (AR/VR) glasses to holographic telepresence. 

\noindent\emph{Digital twins for manufacturing}: Establishing a virtual connection between a digital representation of a complex system or environment and its real-world counterpart.  

\noindent\emph{e-Health for all}: Providing remote medical surgery in areas where doctors and infrastructure are lacking. 

\noindent\emph{Cooperative mobile robots}: Requiring deterministic communication for handling critical motion control information. 

\noindent Table~\ref{tab:UseCases} quantifies the performance requirements for the above use cases. \blue{To approach these latency and reliability requirements,} Wi-Fi is considering a paradigm shift towards introducing more performance determinism. This is not an easy task, since \blue{unlike 3GPP technologies like 5G operating in licensed bands, Wi-Fi operates in unlicensed bands subject to channel access contention and uncontrolled interference. To cope with uncoordinated usage in the unlicensed spectrum, rather than prioritize determinism,} Wi-Fi's medium access control (MAC) was originally designed upon carrier sense multiple access with collision avoidance (CSMA/CA). Evolving from this legacy, Wi-Fi 8 \blue{intends to} pursue determinism through coordination and a \blue{more efficient use of the available spectrum.}

\begin{figure*}
    \centering
    \includegraphics[width=0.85\textwidth]{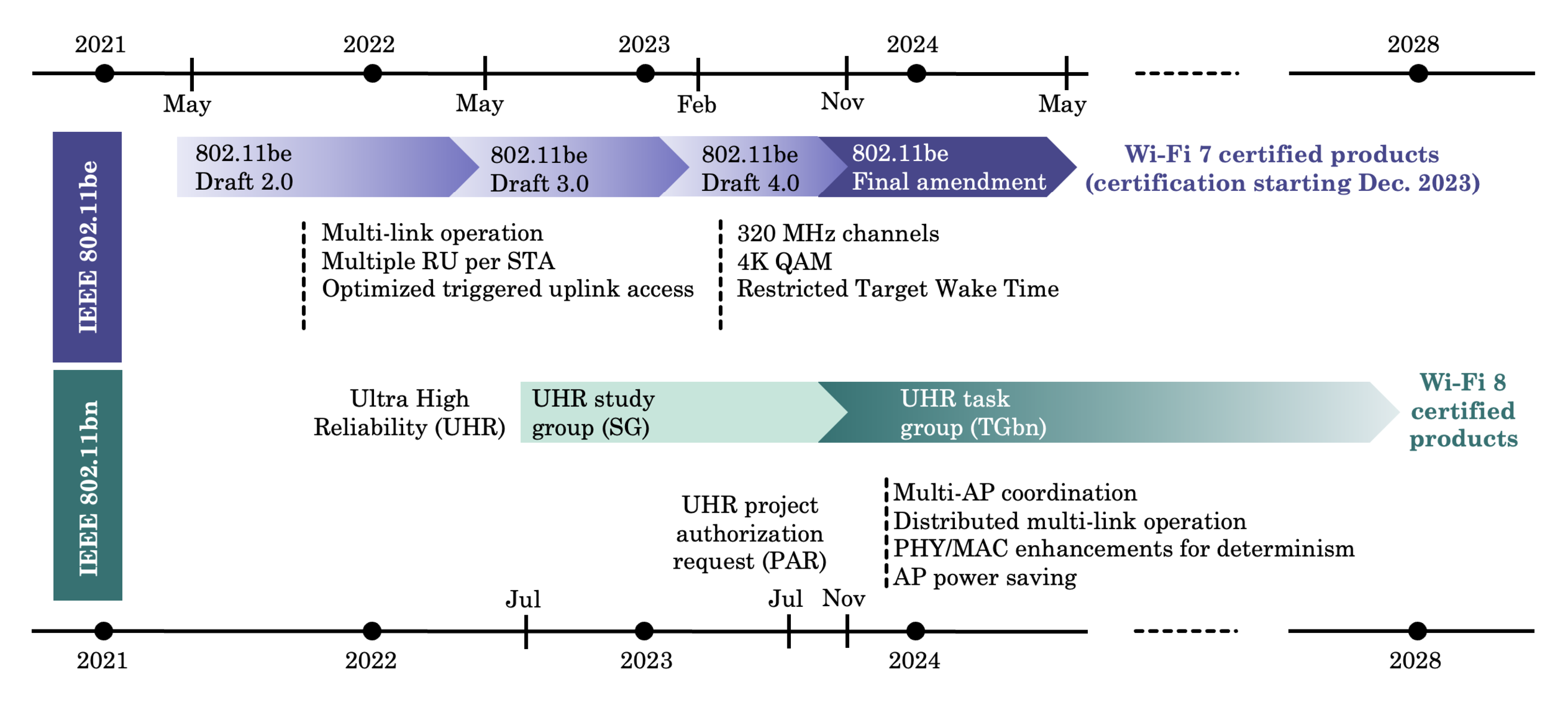}
    \caption{\blue{Current standardization, certification, and commercialization timelines for IEEE 802.11be (top) and IEEE 802.11bn (bottom).}}
    \label{fig:timeline}
\end{figure*}

\blue{\subsection{Novel Standardization Efforts}}
\label{sec:standardization}

\subsubsection*{IEEE 802.11 Real Time Applications (RTA) Topic Interest Group (TIG)}
Back in 2019, the RTA-TIG provided a set of recommendations and guidelines to support low latency and reliability in future Wi-Fi networks \cite{RTA}. Those recommendations have been considered in the Wi-Fi 7 development (e.g., MLO), but they are also influencing the effort towards reliability in Wi-Fi 8, e.g., \blue{via time-sensitive networking (TSN)} integration.

\subsubsection*{IEEE 802.11 AI/ML Topic Interest Group (TIG)} 
Established to explore the application of artificial intelligence (AI) and machine learning (ML) \blue{directly to Wi-Fi protocols. Its aim is to discuss relevant use cases, together with their technical feasibility based on existing mechanisms and expected implementation efforts. These include channel state information (CSI) feedback compression using neural networks, enhanced roaming assisted by AI/ML, deep reinforcement learning-based channel access, and enhanced multi-AP coordination schemes driven by AI/ML \cite{AIMLTIG}.}

\subsubsection*{\blue{IEEE 802.11 Integrated mmWave Study Group (IMMW SG)}} 
\blue{To ensure the long-term evolution of Wi-Fi, next-generation high-end devices could also potentially operate in all three sub-7 GHz bands as well as in the mmWave realm. Indeed, there is a growing interest in better capitalizing on the up to 14 GHz of licensed-exempt spectrum available nearly worldwide in the 60 GHz bands or 5.5 GHz in the 45 GHz band in China, respectively. 
The 60 GHz band is currently used by several incumbent technologies, such as satellite, radio astronomy, and IEEE 802.11ad/ay (WiGig). However, the market adoption of WiGig has been confined to niche applications, and regulatory bodies may consider repurposing the 60 GHz band for other bandwidth-hungry technologies such as 5G and 6G. Against this background, and after initial discussions about extending the UHR scope, it was decided to create a dedicated IMMW SG to pose the basis for the development of a new 802.11 amendment, leveraging PHY/MAC functionalities of the existing Wi-Fi 7 and future Wi-Fi 8 radio interfaces for the sub-7 GHz bands, including channelization and multi-link framework to dynamically operate additional mmWave links.}

\subsubsection*{IEEE 802.11bn UHR} 
\begin{figure*}
    \centering
    \includegraphics[width=0.85\textwidth]{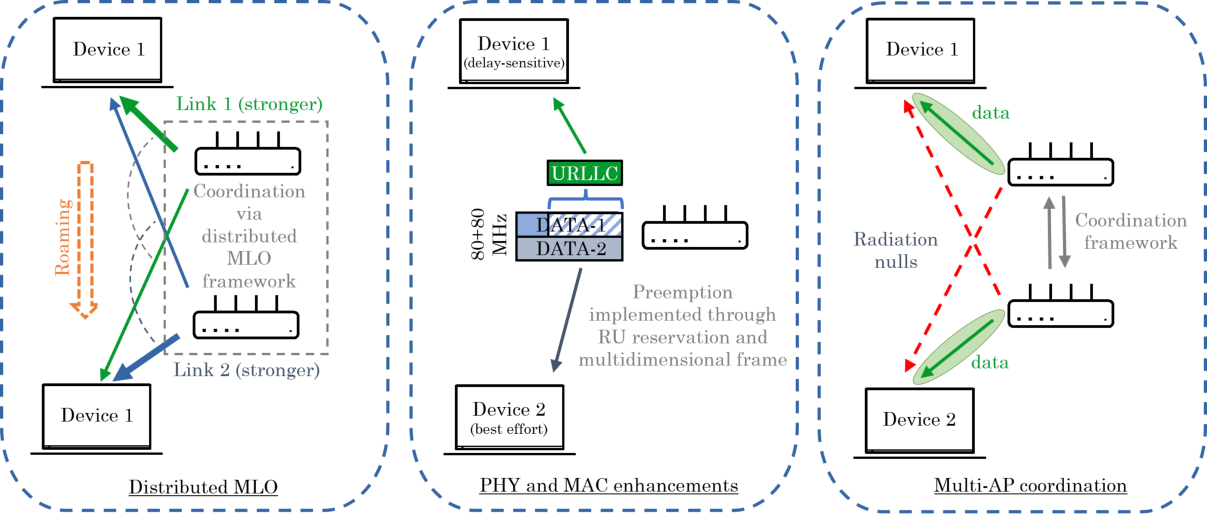}
    \caption{Illustrative examples of the key features being investigated for Wi-Fi\,8.} 
   \label{fig:WiFi8Features}
\end{figure*}

\blue{Fig.~\ref{fig:timeline} summarizes the ongoing IEEE standardization effort for 802.11bn (bottom) alongside the nearly completed 802.11be amendment (top) and its consolidated main features.} 
\blue{The UHR Study Group (SG)} was established in July 2022 to discuss and produce a new Project Authorization Request (PAR) defining the set of objectives, frequency bands, and technologies to be considered beyond 802.11be. The \blue{resulting UHR Task Group (TG) was formed in November 2023}, with a traditional single release standardization cycle that will last until 2028. This activity will define the protocol functionalities of future Wi-Fi\,8 products, mainly focusing on these aspects to be improved with respect to 802.11be \cite{UHRProposedPAR}:
\blue{
\begin{itemize}
\item Increasing throughput by 25\%, as measured at the MAC data service AP. 
\item Reducing by 25\% the 95th percentile latency and by 25\% MAC Protocol Data Unit (MPDU) loss, even in scenarios with mobility and overlapping basic service sets (OBSSs).
\item Improving power saving mechanisms for the AP and enhancing direct peer-to-peer data exchanges.
\end{itemize}
Three main critical aspects impacting reliability in the unlicensed spectrum are being investigated: seamless connectivity, determinism, and controlled worst-case delay. Fig.~\ref{fig:WiFi8Features} depicts examples for each, with their chief opportunities and challenges discussed in the next three sections, respectively.
}

\section{\blue{Seamless Connectivity via Distributed MLO}}

\blue{The multi-link architecture introduced in IEEE 802.11be offers a high degree of flexibility, presenting a clear split between upper (multi-link level) and lower (link level) MAC functionalities, with a multi-link device (MLD) that can be viewed as an entity controlling two or more legacy APs (or STAs), each operating on a single link and co-located on the same hardware \cite{garcia2021ieee,khorov2020current,CheCheDas22}. 
This multi-link framework already allows a multi-link station to switch links with minimal signaling overhead and delay, implicitly enabling seamless transitions between APs under the control of the same MLD entity, and thus allowing for a make-before-break path switch.}

\subsubsection*{Opportunities}
\blue{To improve mobility support, one of the major sources of unreliability in Wi-Fi, 802.11bn has the possibility to extend the just described multi-link architecture towards a distributed framework, where APs under the control of the same MLD entity do not necessary have to be co-located on the same physical hardware. 
This approach creates a distributed virtual cell where a device’s mobility is seamlessly handled by enabling multiple links to be concurrently activated from different distributed APs, thus ensuring that a nomadic device is always connected to at least one link, effectively embedding native roaming support into 802.11bn and significantly improving the connection’s reliability.}

\subsubsection*{Technical challenges} 
\blue{Several critical aspects would need to be addressed to implement distributed multi-link operations in 802.11bn. Primarily, the distributed MLO approach requires coordination and communication among the different distributed APs under the same controlling multi-link instance. In addition, different links would need unique addressing, considering that current 802.11be specification guarantees different identifiers only for links activated within the same physical device.}

\subsubsection*{Potential implementation}
\blue{The coordination among the different distributed APs may be implemented following different approaches. One option is to define a mobility domain where APs, either co-located or not, could be affiliated with an extended MLD entity. Another possible option is to consider a novel overarching logical entity that would provide seamless roaming between links located in two or more 802.11be AP MLDs. In addition, the 802.11bn distributed MLD architecture should define novel, reliable, and sufficiently general interfaces between the coordination entity (e.g., MLD upper MAC) and the coordinated APs (e.g., MLD lower MAC) to allow for the usage of both wired and wireless communications and to ensure interoperability among implementations provided by different vendors.}
\section{\blue{Determinism via PHY and MAC Enhancements}}

\blue{Considering traffic characteristics is crucial in designing low-latency mechanisms. While it would be desirable to handle traffic with predictable arrival patterns by leveraging existing solutions \cite{bankov2020tuning}, the challenge intensifies when accommodating unexpected, event-driven, time-sensitive traffic.}

\subsubsection*{Opportunities} \blue{Upon the arrival of an unexpected high-priority packet at a device, two primary sources of latency may be encountered: the remaining time of another ongoing transmission and the following channel contention procedure for its own transmission. IEEE 802.11bn can tackle both aspects by: (i) extending enhanced distributed channel access (EDCA) with additional priority classes and associated channel access parameters, e.g., for backoff; (ii) extending OFDMA implementation by introducing resource unit (RU) reservations and enabling preemption; and (iii) taking advantage of channel access opportunities in secondary channels when the primary channel is occupied by other transmissions.}

\subsubsection*{Proposed mechanisms} 
\blue{Addressing the latter two opportunities, 802.11bn is contemplating the introduction of two MAC enhancements.
\emph{Resource Reservation and Channel Preemption} could potentially reserve a small RU for low-latency traffic in all transmissions. Coupled with pre-padding, this would enable a node to promptly serve incoming low-latency packets by allocating them to the reserved RU. However, to avoid RU wastage across all transmissions, this RU could also be used for the transmission of actual data, provided that preemption is supported (see Fig.~\ref{fig:WiFi8Features}, second example). Notably, this approach does not impose receiver design changes, but it does require designing a multidimensional PPDU frame. However, if the device aiming to transmit a time-sensitive frame is not the transmission opportunity (TXOP) holder, effective preemption mechanisms must exploit short inter-frame spaces between transmitted PPDUs to seize the channel.} 
\blue{In addition, \emph{Secondary Channel Access (SCA)} may extend the preamble puncturing functionalities in 802.11be by removing dependency on the primary channel and better leveraging transmission opportunities on an idle secondary channel only. The introduction of SCA is anticipated to yield performance gains in scenarios with medium-to-high load without entailing excessive complexity.} 

\section{\blue{Controlled Worst-Case Delay via\\ Multi-AP Coordination}}

Random access procedures \blue{make it difficult} to provide performance guarantees in Wi-Fi, which is why R-TWT was introduced in 802.11be to reduce contention within a basic service set (BSS) \blue{by means of scheduling coordinated service periods}. However, inter-BSS interactions are still governed by contention principles, even if the APs belong to the same administrative domain, making worst-case delays unpredictable. Wi-Fi\,8 is expected to address this issue by introducing multi-AP coordination (MAPC) to achieve greater reliability and prevent channel access contentions, especially in dense and heavily loaded environments.
\blue{
\subsubsection*{Channel state information acquisition}
The implementation of multi-AP coordination mechanisms relies on OBSS CSI, i.e., on estimating the channel for non-associated neighboring devices. 
A certain BSS AP can initiate the OBSS channel sounding procedure through a trigger frame that indicates the IDs of the STAs and AP in the OBSS. The OBSS AP follows by transmitting control frames for sounding (e.g., NDPA and NDP). The OBSS STAs then respond by feeding back the measured channel information to both the BSS AP and the OBSS AP. This procedure may be performed several times to acquire channel state information from multiple OBSS \cite{mentorLG_0854r0}.
Availing of such information can be essential to manage frequency resources, adjust the transmit power, or devise specific beamforming methods so as to avoid OBSS interference. As described in the sequel, each of the different AP coordination schemes may require a different amount of channel state information (e.g., overall signal strength vs. per-antenna small-scale fading estimation) with a very different periodicity. As the overheads incurred may offset the performance gains, efficient CSI acquisition will be key for some of these schemes to be part of Wi-Fi 8.
}
\blue{
\subsubsection*{Protocol upgrades}
New frames} will be necessary for discovering and managing multi-AP groups, sharing channel and buffer state data between APs, and triggering coordinated multi-AP transmissions to minimize inter-BSS collisions and achieve a more efficient and dynamic spectrum usage. 
AP coordination schemes in Wi-Fi\,8 are envisioned to leverage both over-the-air and wired signaling. These schemes will range from basic to advanced, depending on the amount of data that must be exchanged between access points and their implementation complexity. While it is still to be decided what aspects of the coordination mechanism will be specified by the standard and what will be left for implementation, \blue{the main schemes may possibly include some of those described in the remainder of this section.}
%
%
%
\subsection{\blue{Coordinated TDMA/OFDMA}}
\blue{These are} two basic approaches leveraging the time and frequency domain, respectively. In C-TDMA, a TXOP is divided in slots and sequentially allocated to different APs. In C-OFDMA, different portions of the band are allocated to different APs. 

\blue{For instance, with C-OFDMA, an AP that obtains a TXOP is able to share its frequency resources with a set of neighboring APs. The minimum resource unit to be employed is currently under discussion, with smaller units (20\,MHz) offering more flexibility and scheduling gain than larger ones (80\,MHz), but also potentially requiring PHY format changes.}

\blue{On the one hand, C-OFDMA can achieve latency reduction by reducing channel contention. On the other hand, the sharing AP faces computational burden and overhead, as it must first request neighboring APs to report their channel and buffer status, and then schedule and allocate resources accordingly.}


\subsection{\blue{Coordinated Spatial Reuse}}
\blue{In coordinated spatial reuse (C-SR), APs cooperatively control their transmit power, allowing concurrent transmissions and thus increasing the total area throughput.} 
\blue{
\subsubsection*{Opportunities}
This approach, incorporating cooperation, represents an upgrade over status-quo 802.11ax spatial reuse, whereby one AP transmits at maximum power and all others must reduce their power accordingly, sometimes to an extent that does not yield a sufficiently high signal-to-interference-plus-noise ratio (SINR). Instead, coordinating the transmit power among APs allows for the guarantee of an adequate SINR at all receiving STAs and to create extra spatial reuse opportunities. Additionally and unlike C-TDMA/OFDMA, C-SR allows parallel transmissions on the same time/frequency resources, and thus potentially achieves a higher throughput and reduced queuing delay.
}
\blue{
\subsubsection*{Technical challenges}
C-SR requires measuring the receive signal strength information (RSSI) for interfering links in order to compute the appropriate transmit power. However, since the RSSI is relatively static, such information could be acquired via beacon measurements, incurring only limited overhead. Accounting for beamforming in the computation of the RSSI (and thus of the transmit power) may yield better performance but also increase complexity and overhead.}
\blue{
\subsubsection*{Potential implementation}
In a measurement phase, a sharing AP can request intra-BSS STAs to measure and report their RSSI from other APs. 
Once the sharing AP gains access to a TXOP, it collects information from other APs, including which STAs those APs intend to transmit to and their target SINRs. Based on this knowledge, the sharing AP can then calculate the appropriate transmit power for each of the other APs. This information is then communicated via a trigger frame, along with the sharing AP's transmit power, allowing the other APs to set their optimal modulation and coding schemes.
}


\subsection{\blue{Joint Transmission}}
Joint transmission \blue{(JT)} is an advanced approach, also known as \emph{distributed MIMO}, leveraging the spatial domain and involving non-co-located APs that jointly transmit/receive data to/from multiple STAs. 
\blue{
\subsubsection*{Opportunities}
Remarkably, JT turns neighboring APs from potential interferers to servers. This approach has the potential to simultaneously achieve high throughput and low latency, since interference can be suppressed without sacrificing the number of spatial streams. 
}
\blue{
\subsubsection*{Technical challenges}
The success of JT may depend on designing a new distributed CSMA/CA protocol and ensuring tight synchronization in time, frequency, and phase among the cooperating APs. 
Moreover, this feature requires all APs involved to share the data to be transmitted. 
In order to limit the ensuing overhead and prevent an undesired increase in the queueing delay, joint transmission is likely to require an out-of-band backhaul link to connect the APs, e.g., a 10\,Gbps Ethernet cable \cite{mentorSony_1821r1}.
}
\blue{
\subsubsection*{Potential implementation}
A certain AP (AP$_1$) exchanges a coordination request/response with another AP (AP$_2$) to decide whether coordination should be started and which packets would be sent jointly. AP$_1$ then transmits a coordination set to AP$_2$ to start data sharing, e.g., via a wired backhaul. Once data sharing is completed, AP$_1$ sends a coordination trigger to AP$_2$ to start a coordinated transmission, at the end of which both APs receive a block acknowledgment from the receiving STAs. 
Possible solutions to limit the overhead introduced by data sharing could be: (i) completing the data sharing in advance, rather than prior to transmission, whenever possible; (ii) performing wireless packet transmissions to other STAs during wired data sharing to improve efficiency.
}


\subsection{\blue{Coordinated Beamforming}}
\blue{Coordinated beamforming (CBF), also leveraging the spatial domain, is an approach where collaborative APs suppress incoming OBSS interference (see Fig.~\ref{fig:WiFi8Features}, third example).}
\blue{
\subsubsection*{Opportunities}
With CBF, a next-generation multi-antenna AP uses its spatial degrees of freedom not only to multiplex its own STAs but also to place radiation nulls to and from neighboring non-associated STAs. This approach makes the AP and its neighboring STAs mutually \emph{invisible}, avoiding channel access contention, allowing transmissions at full power, and potentially improving worst-case latency as a byproduct. 
}
\blue{
\subsubsection*{Technical challenges}
Unlike JT, CBF does not require joint data processing as each STA transmits/receives data to/from a single AP, therefore not incurring the data sharing overhead and removing the off-band backhauling needs. However, impact of overhead should be carefully considered when defining the CSI acquisition framework. With the size of the antenna arrays expected to grow, 802.11bn should compare the benefit of a more accurate explicit procedure, naturally entailing higher overhead, with an implicit one that trades accuracy for overhead reduction.   
In addition, since the spatial degrees of freedom are limited by the size of the antenna arrays, an appropriate tradeoff between spatial streams carrying data, beamforming gains, and nulling accuracy should be found, along with opportunistic user scheduling during each newly created spatial reuse opportunity.
}
\blue{
\subsubsection*{Potential implementation}
CBF is likely to entail the design of the following key phases:
\begin{itemize}
    \item A control frame exchange between two or more collaborative APs to establish and maintain a coordination set. 
    \item A CSI acquisition phase, for APs to communicate with OBSS STAs and configure space-domain interference suppression. The latter modifies the conventional filter employed for spatial multiplexing---e.g., a zero forcing (ZF) or minimum mean square error (MMSE) precoder---by imposing nulls on a specific channel direction (aiming for complete nulling towards a certain STA) or subspace (aiming for partial nulling towards multiple STAs).
    \item A framework for dynamic nullsteering-based spatial reuse, wherein a donor AP grants a transmission opportunity to an OBSS AP by communicating its served STAs and correspondent interference suppression conditions, i.e. the obligation for the OBSS AP to place nulls towards the STAs served by the donor AP \cite{mentorUnisoc_0855r1}. 
\end{itemize}
}

\blue{While exhibiting a lower implementation complexity than JT, CBF could circumvent channel contention and, under the right circumstances, could deliver a substantial reduction in worst-case latency. Although the potential of CBF has been recently demonstrated for a single link operation \cite{mentorBroadcom_0855r1,garcia2021ieee}, in the following, we conduct a preliminary evaluation of the performance tradeoffs that arise when CBF is paired with MLO, as envisioned in 802.11bn.}

\section{Case Study: Ultra High Reliability in Wi-Fi\,8}

\begin{figure}
    \centering
    \includegraphics[width=0.42\textwidth]{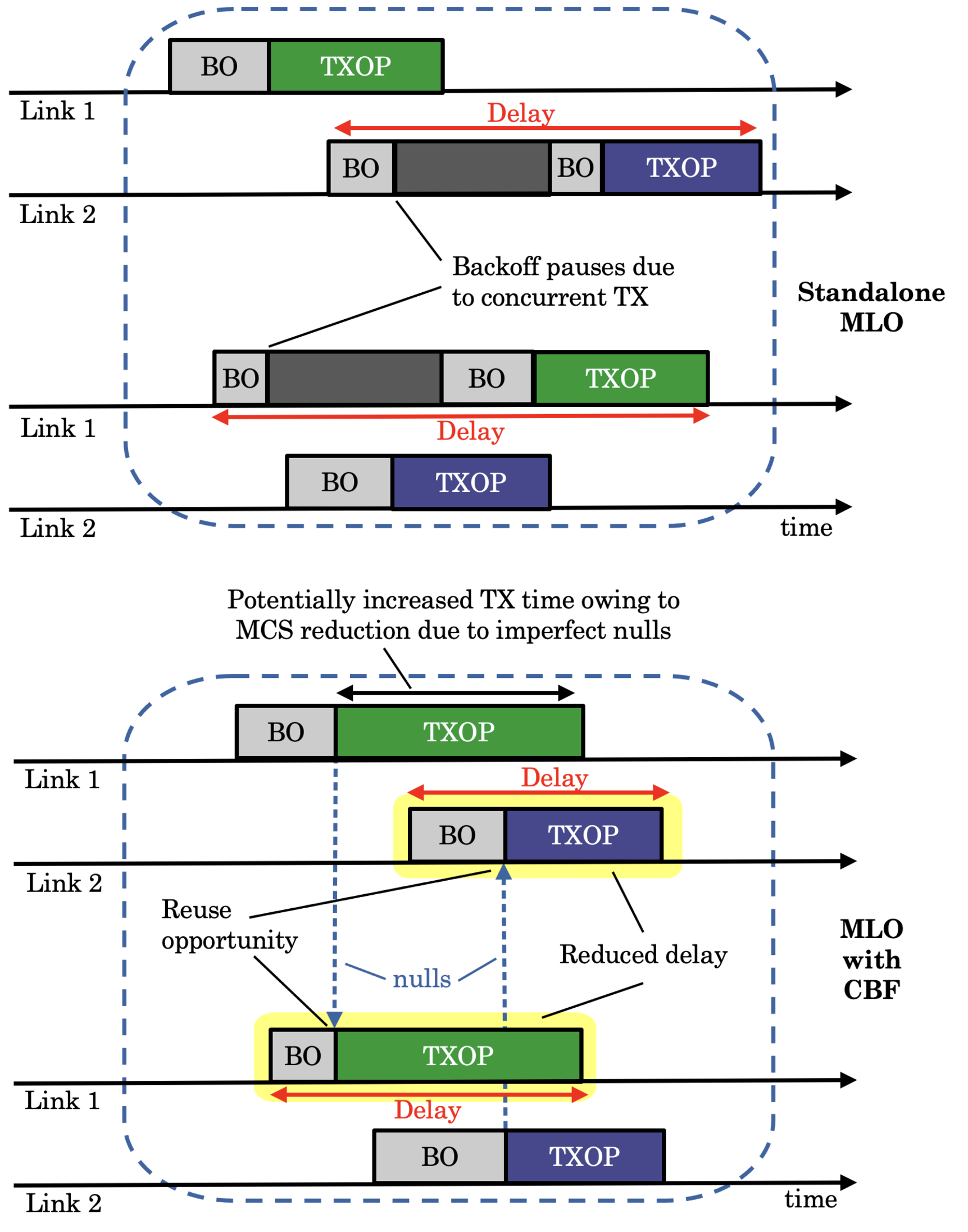}
    \caption{\blue{Illustration of standalone MLO (top) and MLO with CBF (bottom). When CBF is enabled, imperfect nulling may lead to a lower MCS and increased transmission delays. However, this may be more than compensated for by the increased spatial reuse.}}
    \label{fig:MLO+CBF}
\end{figure}

\blue{To assess the potential benefits introduced by MAPC---and specifically CBF---we consider a state-of-the-art Wi-Fi 7 MLO-capable network} that consists of two overlapping BSSs similar to the rightmost scenario shown in Fig.~\ref{fig:WiFi8Features}. Each BSS includes a single AP, equipped with four antennas, and a single associated STA, equipped with two antennas. The two BSSs support \blue{multi-radio MLO (EMLMR) \cite{carrascosa2022understanding}}, operate on the same two 160 MHz links in the 6 GHz band, and implement CBF. Each AP transmits two spatial streams to their respective STAs, using their two remaining spatial degrees of freedom to create radiation nulls towards the other BSS when CBF is enabled.
A 2\,Gbps traffic stream is active on each AP, corresponding to a 120 frames-per-second holographic video stream with ON/OFF activity periods of 4.15\,ms each. 
All latency values refer exclusively to the AP-STA delay and other Wi-Fi 7 and potential Wi-Fi 8 features are not implemented to isolate and highlight the gains provided by CBF. 
\blue{We evaluate CBF performance using variable overheads and nulling accuracy to model the effect of CSI acquisition signaling and its aging.} 
\blue{The same simulator as the one used in \cite{carrascosa2022understanding} is employed and the number of generated packets exceeds 16.5 million.} The full set of simulation parameters is reported in Table~II.

\begin{figure}
    \centering
    \includegraphics[width=0.45\textwidth]{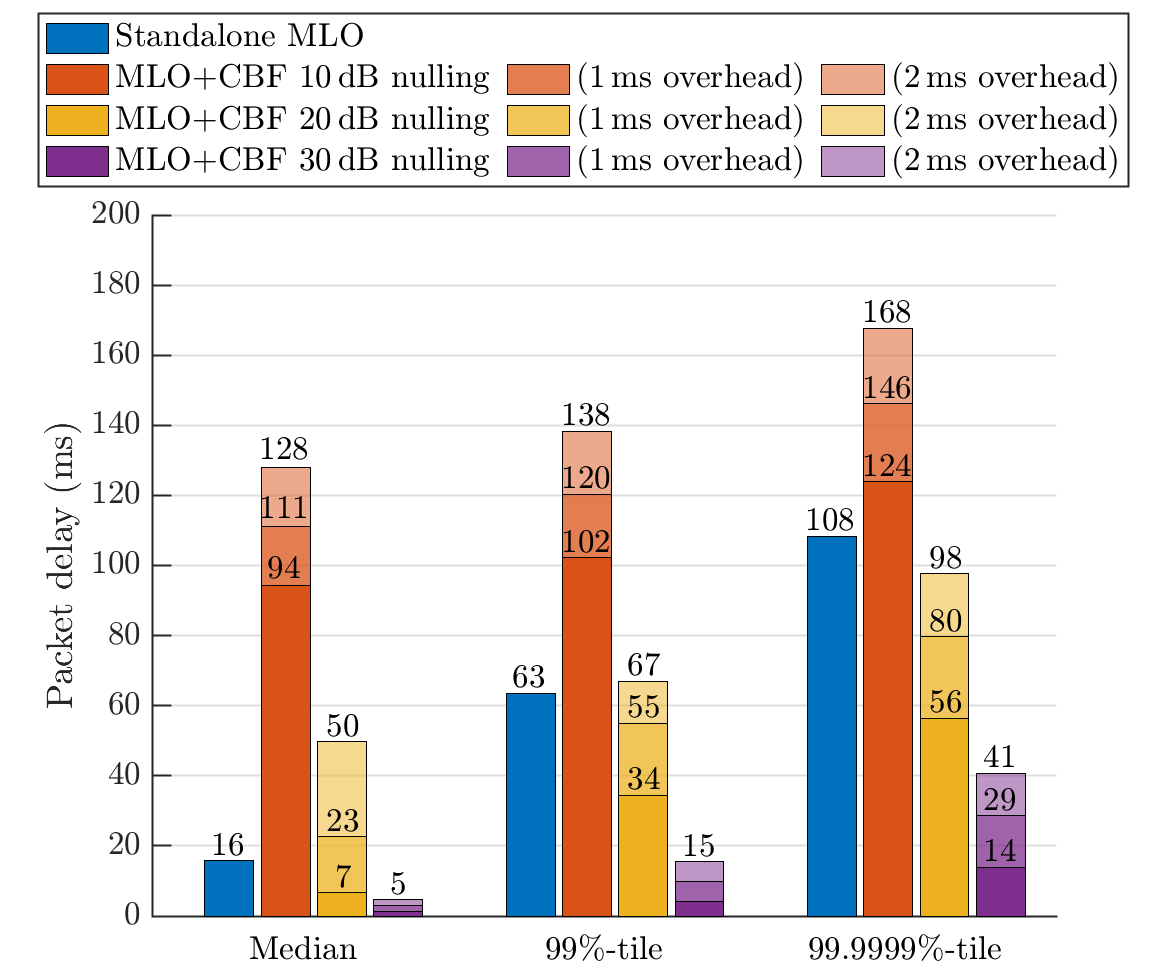}
    \caption{\blue{Delay incurred by standalone MLO and when combining MLO with CBF under a variable nulling accuracy and overhead.}}
    \label{fig:ThroughputAndDelay}
\end{figure}

As illustrated in Fig.~\ref{fig:MLO+CBF}, employing MLO with CBF (bottom) creates additional reuse opportunities and reduces the delay when compared to standalone MLO (top). However, the achievable performance of CBF is related to the accuracy in the null placement. \blue{Fig.~\ref{fig:ThroughputAndDelay} presents the median, 99\%-tile, and 99.9999\%-tile delay values obtained by combining MLO with CBF as the interference suppression accuracy increases from 10 to 30 dB (different colors) and the CSI acquisition overhead increases from 0 (opaque) to 1\,ms (semi-opaque) and 2\,ms (transparent). For comparison, the corresponding performance with standalone MLO is also displayed.} The results show that when nodes contend for the medium with standalone MLO, the 99.9999\% delay exceeds 100 ms. Such performance worsens when combining MLO with CBF at a null accuracy of just 10\,dB, as the benefits of higher spatial reuse are outweighed by the resulting increase in interference and degraded modulation and coding scheme (MCS), down to 16-QAM 3/4. However, the trend reverses when the null accuracy increases to 20\,dB and above, as the MCS reduction incurred is more than compensated for by a lack of contention. An accuracy of 30 dB or more allows for the highest MCS (4096-QAM 5/6) and nearly an order of magnitude reduction in the 99.9999\%-tile delay. \blue{The figure also indicates that CSI acquisition overheads of 1\,ms and 2\,ms can nearly outweigh a CBF nulling capability of 20\,dB.}

\blue{The presented results demonstrate that even with Wi-Fi 7 features such as 4096-QAM and MLO, meeting low delay requirements with ultra-high reliability can be challenging in dense scenarios where all links exhibit high contention levels.} CBF can address this issue and help Wi-Fi 8 cope with use cases requiring reliably high throughput and low latency, such as future immersive holographic communications.


\begin{table}
\centering
\caption{System-level simulation parameters for the case study.}
\label{table:parameters}
\def\arraystretch{1.2}
\colorbox{BackgroundGray}{
\begin{tabulary}{\columnwidth}{ |p{2.6cm} | p{5.0cm} | }
\hline
\rowcolor{BackgroundLightBlue}
  \textbf{Deployment \& traffic}	&  \\ \hline
  AP locations			& (5\,m, 10\,m) and (10\,m, 10\,m) \\ \hline
  STA locations			& (5\,m, 12.5\,m) and (10\,m, 12.5\,m) \\ \hline
  Traffic arrivals			& On/Off, both periods exponentially distributed with mean 4.15\,ms (2\,Gbps per AP) \\ \hline
\rowcolor{BackgroundLightBlue}
  \textbf{MAC}	&  \\ \hline
  Buffer size, TXOP			& 10,240 packets, 5.484\,ms max.  \\ \hline
  Frame aggregation			& 1024 frames max., 10\% packet error rate \\ \hline
  MLO mode			& STR EMLMR \\ \hline
\rowcolor{BackgroundLightBlue}
  \textbf{PHY}	&  \\ \hline
  Frequency band			& 6\,GHz band with two 160\,MHz channels \\ \hline
  Channel model			& IEEE 802.11ax (indoor residential) \\ \hline
  Transmit power, noise			& 20\,dBm, -174\,dBm/Hz spectral density  \\ \hline
  MCS selection			& SINR-based, highest: 4096-QAM 5/6 \\ \hline
  Number of antennas			& 4 per AP and 2 per STA \\ \hline
  CBF nulling			& Imperfect, ranging from 10\,dB to 30\,dB \\ \hline
\end{tabulary}
}
\vspace{-0.5cm}
\end{table}



\section{Conclusion}

In this paper, we discussed why and how the Wi-Fi community is betting on ultra-high reliability with 802.11bn. \blue{We began by outlining the upcoming use cases and} sharing updates on standardization, certification, and spectrum regulation. We then explored the disruptive innovations that Wi-Fi\,8 is likely to bring to meet new requirements previously out of reach in the unlicensed spectrum. We further presented a novel system-level study, demonstrating how Wi-Fi\,8 could achieve ultra-high reliability \blue{through the joint interworking of multi-link operation and spatial-domain multi-AP coordination.} 
\section*{\blue{Acknowledgments}}

\blue{\noindent This work was in part supported by grants PID2021-123995NB-I00, PRE2019-088690, PID2021-123999OB-I00, CEX2021-001195-M, the UPF-Fractus Chair, and the UNICO 5G I+D SORUS project.} \blue{The constructive feedback from the Editor, Prof. Gunes Karabulut Kurt, and the anonymous Reviewers is gratefully acknowledged.}


%
%
\bibliographystyle{IEEEtran}
\bibliography{journalAbbreviations, bibl}
\section*{Biographies}
\small

\noindent
\textbf{Lorenzo Galati Giordano} (SM’20) is a Senior Research Engineer at Nokia Bell Labs. He has more than 15 years of experience resulting in tens of commercial patents, publications, and standard contributions. 

\vspace{0.2cm}
\noindent
\textbf{Giovanni Geraci} (SM’19) is a Principal Research Scientist with Telef\'{o}nica Research, an Associate Professor and the Head of Telecommunications Engineering at Univ. Pompeu Fabra in Barcelona, and serves as a Distinguished Lecturer for IEEE ComSoc and IEE VTS.

\vspace{0.2cm}
\noindent
\textbf{Marc Carrascosa} is currently a Ph.D. candidate in the Wireless Networking group at Universitat Pompeu Fabra (UPF). His research interests are related to performance optimization in wireless networks.

\vspace{0.2cm}
\noindent
\textbf{Boris Bellalta} (SM’13) is a Full Professor at Universitat Pompeu Fabra (UPF). His research interests are in wireless networks, with emphasis on Wi-Fi and machine learning-based adaptive systems.
%
\end{document}